\documentclass{ieeeaccess}
\usepackage{cite}
\usepackage{amsmath,amssymb,amsfonts}
\usepackage{algorithmic}
\usepackage{graphicx}
\usepackage{textcomp}
\def\BibTeX{{\rm B\kern-.05em{\sc i\kern-.025em b}\kern-.08em
    T\kern-.1667em\lower.7ex\hbox{E}\kern-.125emX}}
\begin{document}
\history{Date of publication xxxx 00, 0000, date of current version xxxx 00, 0000.}
\doi{10.1109/ACCESS.2017.DOI}

\title{Mitigating domain shift in AI-based tuberculosis screening with unsupervised domain adaptation}
\author{\uppercase{Nishanjan Ravin\authorrefmark{1}, Sourajit Saha\authorrefmark{1},
Alan Schweitzer\authorrefmark{2}, Ameena Elahi\authorrefmark{2}, Farouk Dako\authorrefmark{2},  Daniel Mollura\authorrefmark{2}, David Chapman\authorrefmark{1}}.}
\address[1]{Department of Computer Science and Electrical Engineering, University of Maryland Baltimore County, 1000 Hilltop Circle, Baltimore, MD, 21250 (email: dchapm2@umbc.edu)}
\address[2]{Rad-AID International nonprofit, 8004 Ellingson Drive, Chevy Chase, MD 20815 (e-mail: info@rad-aid.org)}
\tfootnote{``This work was supported in part by the NSF Center for Advanced Real-time Analytics Grant 1747724 and in part by Google Foundation.''}
 
\markboth
{N. Ravin \headeretal: Mitigating domain shift in AI-basedtuberculosis screening withunsupervised domain adaptation}
{N. Ravin \headeretal: Mitigating domain shift in AI-basedtuberculosis screening withunsupervised domain adaptation}

\corresp{Corresponding author: David Chapman (e-mail: dchapm2@umbc.edu).}

\begin{abstract}
We demonstrate that Domain Invariant Feature Learning (DIFL) can improve the out-of-domain generalizability of a deep learning Tuberculosis screening algorithm.  It is well known that state of the art deep learning algorithms often have difficulty generalizing to unseen data distributions due to "domain shift".  In the context of medical imaging, this could lead to unintended biases such as the inability to generalize from one patient population to another.  We analyze the performance of a ResNet-50 classifier for the purposes of Tuberculosis screening using the four most popular public datasets with geographically diverse sources of imagery.  We show that without domain adaptation, ResNet-50 has difficulty in generalizing between imaging distributions from a number of public Tuberculosis screening datasets with imagery from geographically distributed regions.  However, with the incorporation of DIFL, the out-of-domain performance is greatly enhanced.  Analysis criteria includes a comparison of accuracy, sensitivity, specificity and AUC over both the baseline, as well as the DIFL enhanced algorithms.  We conclude that DIFL improves generalizability of Tuberculosis screening while maintaining acceptable accuracy over the source domain imagery when applied across a variety of public datasets.
\end{abstract}


\titlepgskip=-15pt
\maketitle
\section{Introduction}
\label{sec:introduction}
\PARstart{G}{ENERALIZABILITY} beyond the source domain is an important and difficult challenge for machine learning. In medical imaging, it can have a major impact on \textit{clinical trustworthiness}. This is because it is not known whether a deep learning algorithm trained on patients from one population will generalize to another without extensive \textit{out-of-domain} testing.  This \textit{out-of-domain} testing however is rarely performed in Machine Learning (ML) literature, and instead \textit{in-domain} testing through \textit{cross validation} is much more common, yet inadequate because it does not measure the negative impact of \textit{domain shift} on \textit{real-world} performance.  In the words of Thrall et al. \cite{Thrall_2018}, \textit{"Tolerances of using AI programs in imaging between different patient populations is not yet known.  Failure to recognize that a program is not generalizable, for example from adults to children or between different ethnic groups could lead to incorrect results."} Machine Learning algorithms \cite{zandehshahvar2021toward, kamran2019optic, koziarski2021diagset} have elevated the efficacy of Computerized Aided Diagnosis (CAD) across different classes of medical imagery such as Tomography, X-ray. However, these algorithms are rarely tested for cross domain validation.

\subsection{Cross Validation vs Real World}

The \textit{real-world} performance of machine learning algorithms is universally lower than the reported \textit{cross validated} performance due to \textit{domain shift}.  But how much lower?  Although there is potential for minor degradation, there have also been reported instances of machine learning algorithms including \textit{deep learning} where performance drops to surprisingly low levels when tested \textit{out-of-domain}.  A canonical example is the difficulty of deep learning networks, trained using the MNIST digits dataset, to accurately predict the USPS zip-code digits dataset without retraining or Domain Adaptation \textbf{(cite)}. It is easy for a researcher to erroneously believe that a deep learning algorithm trained to predict MNIST digits could be applied whole-cloth to handwritten digit recognition tasks in the wild such as parsing of postal zip codes or handwritten checks.  But this is often not the case, and most concerning is that this degradation is typically not measured.

Fortunately, many recent works have shown that \textit{Unsupervised Domain Adaptation} (UDA) enabled algorithms trained on MNIST can generalize not only to USPS, but even to the more difficult Street View House Numbers (SVHN) which are typically printed and not handwritten. Furthermore, UDA does not require labels of the target dataset, and can be performed using only unlabeled imagery from the target domain. As such, UDA holds promise as a potential means to overcome the \textit{out-of-domain} degradation that might occur due to \textit{domain shift} so long as representative unlabeled images can be obtained.  

%This is done by first producing generalized features between the different datasets, and then using these generalized features for the classification task at hand. Hence, by using generalized features, the model can be used for identical classification tasks on other datasets. 

\subsection{Cross Validation in AI for Radiology}

Medical imaging datasets contain a unique distribution of meta-attributes including patient demographics, disease presentations, imaging procedures, labeling criteria, scanner equipment and settings. It is possible for a machine learning or a deep learning algorithm to overfit \cite{roelofs2019meta} these criteria and thereby have difficulty in generalizing to imagery in a new hospital institution or clinic, even if it appears to work well in \textit{cross validation} testing, or if it works well in a source institution for which the training data is well-representative.

Nevertheless, the extent of \textit{domain shift} has not been adequately studied in the context of AI for medical imaging. For the purposes of Tuberculosis (TB) screening in particular, patients across different countries may exhibit different disease presentations.  Disease presentation may vary due to the particular strain of the disease that is prevalent in the region but may also vary by the severity of the disease.  For example, in an institution where patients are screened when they present severe symptoms, one might expect that active TB cases may exhibit more advanced pulmonary infection than if patients are screened upon exhibiting milder symptoms.  Furthermore, variation in the x-ray equipment, settings, and patient positioning may affect the performance of deep learning algorithms if this variation is not well represented in the training data.  Changes in image resolution, scanning and compression artifacts may also contribute to \textit{domain shift}.  Finally, inadequate diversity of the patient population demographics may lead to variation in performance and effectiveness across sensitive demographic groups, including potentially biased AI performance between males and females, adults and children, and/or racial/ethnic diversity.

%Perhaps most concerning is that \textit{domain shift} often goes completely unmeasured in published works based on machine learning.  Most published ML algorithms are evaluated only through cross validation.  But cross validation does not measure domain shift, because both the train and test samples are ultimately split from the same distribution. The real-world does not perfectly aligned with the test dataset, and as such the true performance of any ML algorithm will be lower than the reported \textit{cross validated} accuracy.  But how much lower?  Does domain shift cause a minor degradation, or does it cause a ML algorithm to become completely ineffective for an intended use case in medical imaging, and/or potentially biased in ways that may have negative socioeconomic implications.

\subsection{Contributions and Takeaways}

To the best of our knowledge, no prior published work has evaluated nor reported on severity of domain shift, of CNN algorithms across public TB screening datasets.  Furthermore, no prior published work has attempted to reduce this domain shift through Unsupervised Domain Adaptation.  As such, the main contributions of this work are as follows

\begin{itemize}
  \item We measure the severity of \textit{domain shift} for a strong baseline Convolutional Neural Network (CNN) algorithm for Tuberculosis (TB) screening through \textit{out-of-domain-testing} across the three most prevalent public datasets for TB screening with chest radiographs.
  \item We present a novel algorithm for adversarial Domain Invariant Feature Learning (DIFL) for unsupervised Domain Adaptation (DA) with chest radiograph images.
  \item We evaluate the extent to which (DIFL) can prevent or mitigate domain shift and improve algorithmic performance across out-of-domain TB screening tasks.
  \item We qualitatively evaluate the results across three public datasets, and include the opinion of a Board Certified Radiologist in determining the capabilities and limitations of the approach for hospital institutional use cases.
\end{itemize}

The key takeaways are as follows,
\begin{itemize}
\item Severe degradation of accuracy is observed in a strong baseline TB screening algorithm when measured across the geographically diverse public TB X-ray datasets.
\item Unsupervised DA is capable of significantly improving (reducing) this domain shift for all source and target dataset combinations.
\item The extent of the improvement, although significant for all source/target pairs, was more substantial in certain combinations than others, suggesting that certain source datasets are better suited for generalizability with UDA.
\end{itemize}

Finally a discussion of the limitations of this study is included at the end of this manuscript, including the discussion of a hypothesis that \textit{disease presentation} was a major factor in the ability for UDA to enable the CNN algorithm to generalize from one patient population to another.  Furthermore, we discuss limitations of the Adversarial Domain Adaptation using Domain Invariant Feature Learning and propose a strategy for future work to overcome these challenges.

\section{Tuberculosis Screening}
Tuberculosis (TB) is a contagious bacterial infection of the lungs which is widespread globally, thus affecting an estimated 25\% of the world's population. 90\% of those infected will never have symptoms, but 10\% will progress to active TB which frequently causes severe coughing, chest pain, and pulmonary scarring. TB is known to develop more frequently in populations of developing countries, with higher rates of infection in Low and Middle Income Countries (LMIC) than High Income Countries (HIC). In particular, many countries in the African and South American continents have high prevalence of TB infections.

It is desirable to train a deep learning algorithm for TB screening that can generalize to populations from around the world. However, due to the aforementioned problem of domain shift in patient populations across the world, this has proved to be a challenging endeavor. In this paper, the impact of domain shift on TB screening algorithms is explored, by utilizing three major public datasets that are available for the identical task of TB screening from chest X-ray images. By training AI models using regular deep learning algorithms on one dataset and cross-testing them on the other datasets, we can evaluate the presence of bias (i.e. the inability to generalize to other domains) in these AI models that arise due to domain shift.

%Three out of the four public datasets are regional datasets, with chest X-ray images originating wholly from a singular medical institution in Shenzhen (China), New Delhi (India) and Montgomery County (USA). The last public dataset, referred to as the TBX11K dataset, is a much more extensive dataset, containing a large number of chest X-ray images taken from several medical institutions around the world. As such, one might expect that a deep learning model trained using the TBX11K dataset would be able to perform significantly well on identical TB screening tasks using out-of-domain (in this case, non-TBX11K) data, such as the three regional datasets. However, initial testing has revealed that even state-of-the-art deep learning algorithms, such as the ResNet50 model, when trained only on the TBX11K data, achieves greatly reduced out-of-domain accuracy when tested on the other datasets. This problem is attributed to the presence of domain shift in the data, which leads to the TB screening model being unable to generalize well on data outside of the domain that it was trained on.

Three geographically distributed public datasets for TB screening are used as part of this analysis.  Each dataset originates wholly from a singular medical institution, but each of these medical institutions is located in a different country.  As such, for the remainder of this manuscript we refer to each dataset by the origin country.  The first dataset originates from a medical institution in Shenzhen (China) \cite{Jaeger_2014}.  The second originates from an institution in New Delhi (India) \cite{Chauhan_2014}, and the third from an institution in Montgomery County Maryland (USA) \cite{Jaeger_2014}.

The availability of geographically distributed data sources for an identical classification screening task makes this problem an ideal candidate for exploring the possibility of evaluating unsupervised DIFL methods in the context of medical imaging. As mentioned earlier, DIFL methods have found success in relatively simple cross-domain computer vision tasks such as handwritten digit classification and object recognition. However, there is relatively little prior work of employing DIFL methods for relatively complex tasks and particularly in the medical imaging domain. TB screening from chest X-ray images is an important problem as TB is a very common disease in many countries internationally. Furthermore, country-to-country variation in imagery may affect classification performance, but the impact of such \textit{domain shift} on AI classification has not been adequately measured and attempts to reduce this \textit{domain shift} through unsupervised DIFL have not been previously tested.
% As such, this paper sets out to explore the viability of improving deep learning algorithms in the medical field through implementing DIFL methods, by specifically delving into the task of screening TB from chest X-ray images. 

\section{Related Work}

While there has been interest and developments in the field of domain adaptation recently, it is still an area of machine learning that is yet to be fully explored. Domain adaptation has found its place in a wide variety of applications, ranging from being a part of a broader transfer learning routine \cite{Patricia_2014, Ghafoorian_2017, Kouw_2018, Liu_2019}  explicitly modelling the transformation between two or more domains \cite{Zhou_2014, Li_2018} and even in data augmentation \cite{Volpi_2018, Zhang_2019}. With a wide variety of flavors to choose from, such as supervised \cite{Daume_2009, Saha_2011, Motiian_2017} semi-supervised \cite{Daume_2010, Yao_2015, Saito_2019}  and unsupervised \cite{Baktashmotlagh_2013, Ganin_2015, Long_2016}, domain adaptation has been increasingly incorporated for an increasing number of tasks, such as handwritten digit classification \cite{Bousmalis_2017, Murez_2018}, object recognition in images acquired in different conditions \cite{Chopra_2013, Murez_2018}, 3D pose estimation \cite{Cao_2019}, and a variety of other tasks. 

However, the application of domain adaptation has been relatively limited to slightly simpler tasks, and work done in the field of domain adaptation in the context of medical imaging has been relatively limited thus far. While there has been a slight uptick in utilizing domain adaptation techniques for tasks relating to medical imaging as of recent, unsupervised domain adaptation remains relatively unexplored in the context of medical imaging. Logically, unsupervised domain adaptation proves to be the most difficult to achieve in comparison to supervised/semi-supervised, in large part due to the absence of classification labels in the target domain(s). Given the relatively complex nature of machine learning in the field of medical imaging, i.e. the distribution of data is not as easily learnt as with other simpler tasks such as handwritten digit classification, a large majority of domain adaptation work done in the context of medical imaging has shied away from unsupervised domain adaptation. 

Perhaps the closest recent work to ours is the publication of Zhou et al. \cite{Zhou_2021}, which explores the use of semi-supervised domain adaptation methods to improve the accuracy of the label classification across different domains for the task of predicting Covid-19 from chest X-ray images. In their proposed Semi-supervised Open Set Domain Adversarial (SODA) network, they utilize an adversarial semi-supervised method of training, such that the SODA network will be able to learn features that are adaptable to the target domain with relatively high accuracy on both the source and target domains. However, this approach assumes the availability of labeled data (at least in limited quantities) on the target domain, which may not be feasible in all real-world use cases. 

Madani et al. \cite{Madani_2018}, very similarly, looks at the potential of semi-supervised domain adaptation methods to increase the accuracy of label classification on different domains for the task of predicting cardiac abnormalities from chest X-ray images, with the added focus of generating synthetic X-ray images to solve the problem of data scarcity in the field of chest X-ray images. As with the previous paper, this approach also assumes the availability of partially labeled data being available in the target domain. 

While both rely on the process of domain invariant feature learning (DIFL) methods to increase the generalizability factor of the classification model across various domains, utilizing a semi-supervised approach assumes the availability of partially labelled data on all the potential target domains. In reality, such labelled data might not be available in the field of medical imaging (mainly pertaining to chest X-ray images), due to difficulty in obtaining \textit{gold-standard} ground truths for the chest X-ray images. As such, unlabelled data is more easily obtainable, which raises the question of whether unsupervised domain adaptation methods would be as effective as semi-supervised domain adaptation methods for the same/similar tasks.

In this paper, we seek to investigate this hypothesis, i.e. how effective is unsupervised domain adaptation in reducing algorithmic bias due to \textit{domain shift}, as measured by the performance difference of a common CNN algorithm ResNet50 as trained with imagery from one institution, but tested using imagery from another institution.  Existing research has not yet looked into this ambitious translational task of unsupervised domain adaptation as a means of algorithmic tuning to enhance the ability of an algorithm to generalize and maintain accuracy when applied to imagery from a different institution on which the algorithms were initially trained.

\section{Methodology}

\begin{figure}
  \includegraphics[width=3.35in]{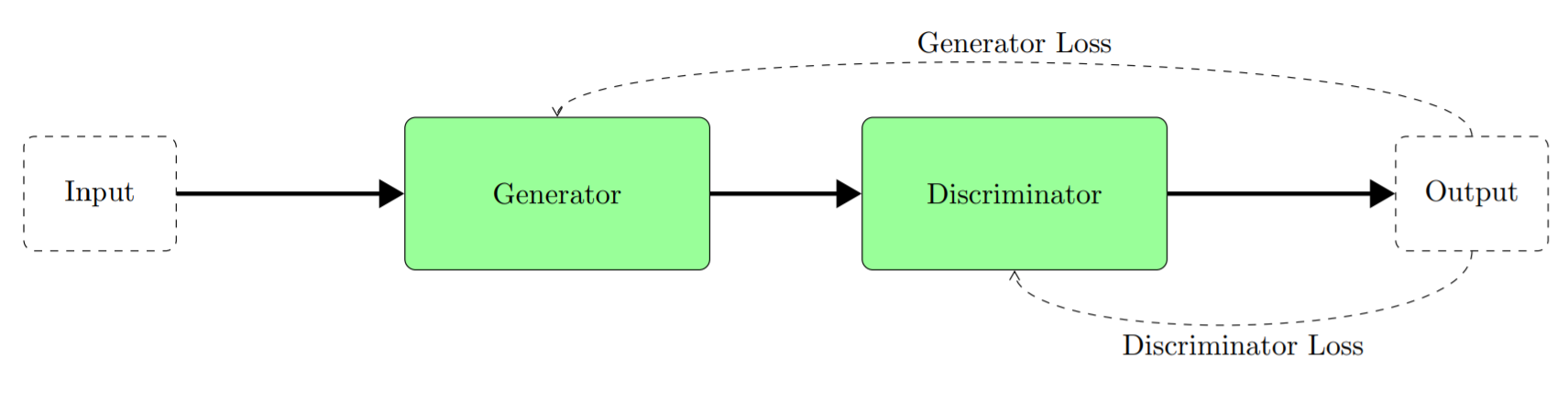}
  \caption{Simplified GAN Architecture.}
  \label{fig1}
\end{figure}

\begin{figure*}
  \includegraphics[width=\textwidth]{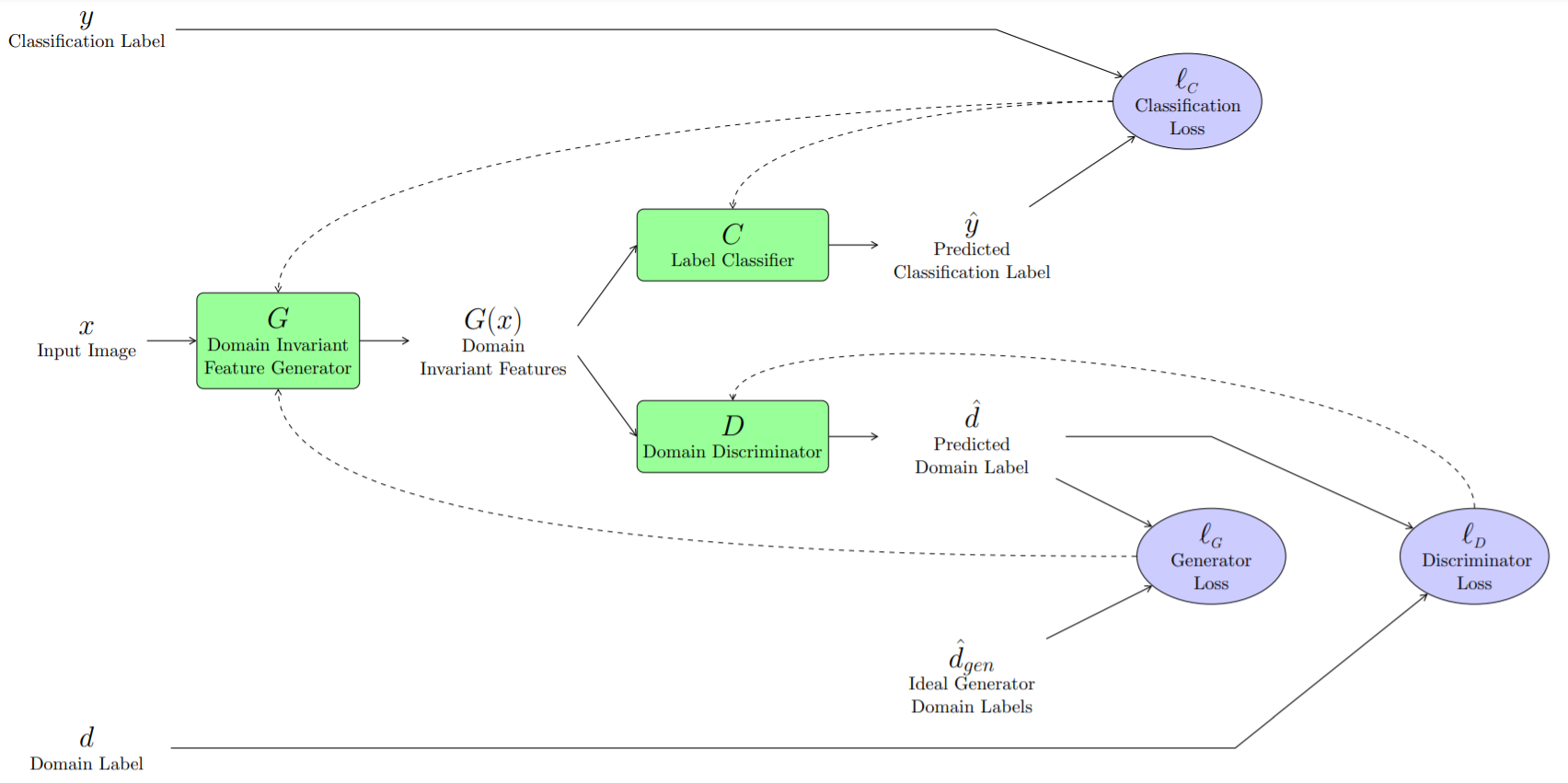}
  \caption{Detailed Overview of the DIFL Model.}
  \label{fig2}
\end{figure*}

%In this section, the developed a Domain Invariant Feature Learning (DIFL) model that will be explained in detail.

%The implemented DIFL model draws inspiration from the Generative Adversarial Network (GAN) architecture, as discussed in \cite{Goodfellow_2014}. The two main components of a basic GAN architecture include the generator network, and the discriminator network. These two networks have opposing goals, with the generator trying to manipulate the input data in a way such that the discriminator fails to successfully categorize the output of the generator into the possible classes, i.e. the generator attempts to “fool” the discriminator, while the discriminator attempts to perform accurate classification. As such, both of these networks have custom loss functions that are adjusted according to the exact function that they serve. A simplified example of the GAN architecture is shown in Figure \ref{fig:simple}. 

In this section, we delineate our proposed unsupervised Domain Invariant Feature Learning (DIFL) model in detail.

The implemented DIFL model draws inspiration from the Generative Adversarial Network (GAN) architecture, as discussed in \cite{Goodfellow_2014}. The two main components of a basic GAN architecture include the generator network, and the discriminator network. These two networks have opposing goals, with the generator trying to manipulate the input data in a way such that the discriminator fails to successfully categorize the output of the generator into the possible classes, i.e. the generator attempts to fool the discriminator, while the discriminator attempts to perform accurate classification. As such, both of these networks have custom loss functions that are adjusted according to the exact function that they serve. A simplified example of the GAN architecture is shown in Figure ~\ref{fig1}. 

In the following discussion, an input image is represented by the variable $x\in X$ where $X$ is the set of all input images, and their corresponding classification labels are defined by the variable $y \in Y$, where $Y$ is the set of all classification labels. The set of images and classification labels are also subdivided into source and target domains, which are denoted by the subscript $S$ and the subscript $T$ respectively. As such, we use the variables $x_S$ to denote images from the set of all source domain images $X_S$, where $x_S \in X_S \subseteq X$, and likewise we use the variable $x_T$ to denote images from the set of all target domain images $X_T$, where $x_T \in X_T \subseteq X$. Similarly, we define the variables $y_S$ and $y_T$ to denote the classification labels from the set of all source classification labels $Y_S$ and the set of all target classification labels $Y_T$ respectively, where $y_S \in Y_S \subseteq Y$ and $y_T \in Y_T \subseteq Y$. However, it is to be noted that the target classification labels $Y_T$ are unobserved, and hence they are not used for training the DIFL model. Hence, the proposed method in this paper is called unsupervised DIFL. 

Additionally, to differentiate between the source and the target domains, a domain label is added for each image $x \in X$. The domain labels are denoted by the variable $d_D$, where $D$ is the set of all domain labels. The variable $d_S$ denotes domain labels taken from the set of all domain labels $D_S$, such that $D_S \in D_S \subseteq D$, and the variable $d_T$ denotes domain labels taken from the set of all domain labels $D_T$, such that $d \in D_T \subseteq D$. Domain labels are used to differentiate the possible domains that any particular image $x$ could be taken from. Thus, for the purposes of this paper, the source domain labels $d_S \in D_S$ for all source images are set to the value of $0$, while the target domain labels $d_T \in D_T$ for all target images are set to the value of $1$. 

A detailed overview of the DIFL model architecture is shown in Figure ~\ref{fig2}, and explained in the following discussion. The DIFL model, and its training process is explained in the following sections. 

\subsection{Training the DIFL Model}

The ultimate aim of the DIFL model is to produce an adequate classification model for the target domain, while using only unlabeled images in the target domain.  In order to accomplish this it is necessary to learn a generalized feature representations of the images $x \in X$, i.e. from both the source and target domains, while also performing successful label classification on the source images $x_S \in X_S$. This overall task is accomplished by training 3 separate neural networks simultaneously: a label classifier network, indicated by $C$, a domain invariant feature generator network, indicated by G, and a domain discriminator network, indicated by $D$. 

In the ideal scenario, the DIFL model would be able to learn and produce perfectly generalizable features of the images $x \in X$, in which case the label classifier network, which has been trained to correctly classify the generalized features of the source images $x_S \in X_S$, can also be utilized to make accurate classifications on the generalized features of the unlabeled target domain images $x_T \in X_T$, as the task is unchanged between the source and target domains. Thus ideally DIFL model will be able to perform equivalently well on both source and out-of-domain data, which is not feasible to be achieved by conventional classification models that are trained using data from a singular domain.

Training the DIFL model involves training all 3 networks G, C and D, where each network makes use of a custom adversarial loss function. Broadly, the training process of the DIFL model can be subdivided into two major steps: a label classification step, and a domain invariance step. The DIFL model is trained by conducting these aforementioned steps simultaneously until convergence of the DIFL model is observed. 

\subsection{Label Classification Step}

In the label classification step, image and classification label tuples $(x_S, y_S)$ from the source domain are used to train a part of the DIFL model. The input images $x_S$ are passed through the domain invariant feature generator network G to produce $G(x_S)$, the \textit{domain invariant features} of the input images $x_S$. These features are then passed through the label classifier network $C$, to obtain $C(G(x_S))$, which are the predicted classification labels of the images $x_S$. We define a variable $\hat{y}_S$ to represent the predicted classification labels of the source images, as follows:

\begin{equation}
y_S=C(G(x_S )) \quad \text{where} \quad x_S \in X_S
\end{equation}

The predicted classification labels $\hat{y}_S$ are then compared with the true classification labels $y_S$, through an appropriate loss function to produce the classification loss, $l_C$. Due to the binary nature of the classification labels, the Binary Cross Entropy loss function is utilized to calculate $l_C$. This can be represented mathematically as follows, where the variable N indicates the total number of images in the batch used for training, and the subscript i is used to denote each individual image:

\begin{equation}
l_C={-{\frac{1}{N}}} \sum_{i=1}^N \; y_s^i \; log(\hat{y}_i)+ (1-y_i) \;log(\hat{y}_i)
\end{equation}

This loss value $l_C$  is used to update the weights of the domain invariant feature generator G and the label classifier C, by differentiating the loss value with respect to the weights of the respective networks, and then multiplying the resulting gradients with a appropriate learning rate before using the resulting values to update the networks themselves. In the following discussions, this particular learning rate is referred to as the classification learning rate, and represented by the variable $\alpha_C$.

\subsection{Domain Invariance Step}

In the domain invariance step, image and domain label tuples $(x,d)$ from both the source and target domains are used to train a part of the DIFL model. The input images x are passed through the domain invariant feature generator network $G$ to produce $G(x)$, the domain invariant features of the input images $x$. These features are then passed through the domain discriminator network $D$, to obtain $D(G(x))$, which are the predicted classification labels of the images $x$. We define a variable $\hat{d}$ to represent the predicted domain labels of the images, as follows:

\begin{equation}
\hat{d} = D(G(x)), \quad \text{where} \quad x \in X
\end{equation}

As the domain invariant feature generator network G and domain discriminator network D are set up in a GAN-like fashion, updating these networks in the domain invariance step is also done in a similar method as with regular GANs. Let us analyze the functions of these two networks in this particular domain invariance step to logically derive the loss functions that should be used to update these two networks. 
Regular GANs utilize the minimax loss function to update the generator and discriminator networks, which was first introduced in (Goodfellow et al., 2014), the same paper that proposed the original GAN structure. The minimax loss function, represented by the value function $V(G,D)$ is as follows:

\begin{equation}
\begin{aligned}
    & min_T \; max_D \\
    & \quad \quad  E_x [ log(D(z_1))] + E_z [log(1-D(G(z_2)))]
\end{aligned}
\end{equation}

While the loss function for the domain invariant feature generator network $G$ and domain discriminator $D$ in the domain invariance step would be similar, it need not necessarily be identical to the above minimax loss function. Let us analyze the functions of these networks in closer detail to set up their respective loss functions.

The domain invariant feature generator network G attempts to produce domain invariant feature representations $G(x)$ of the source and target images $x \in X$, such that the domain discriminator network D is unable to correctly identify which domain the images are taken from. The domain discriminator network $D$ takes $G(x)$ as input and aims to accurately classify them into their appropriate domains, i.e. correctly predict their domain labels.

Thus, the loss function for the domain discriminator network $D$ can be set up in a straightforward manner; the predicted domain labels $\hat{d}$ can be compared together with the actual domain labels $d$ through an appropriate loss function to produce a loss value, represented by the variable $l_D$. For the purposes of this paper, due to the binary nature of the domain labels (the image can only either belong to the source domain or the target domain), the Binary Cross Entropy loss function is used calculate the loss value $l_D$. This can be represented mathematically using the following formula, where the variable N indicates the total number of images in the batch used for training, and the subscript i is used to denote each individual image:

\begin{equation}
    l_D = -{\frac{1}{N}} \sum_i=1^N \; d_i \; log(d_i) + (1-d_i) log(1 - \hat{d}^i)
\end{equation}

Setting up the loss function for the domain invariant feature generator network $G$ requires us to examine its purpose in closer detail. Let us first define the term domain invariant. We can say that the feature representations of a particular image are domain invariant, i.e. the feature representations are highly generalized, when the domain discriminator $D$ is unable to distinguish which domain the image originates from. Mathematically, this statement can be interpreted in the following manner: the domain discriminator $D$ assigns an equal probability for the image to be from the source and the target domain. 

Thus, the domain invariant feature generator network’s goal would be for the domain discriminator network’s output, the predicted domain labels $\hat{d}$, to indicate a probability of $0.5$ for both the source and target domains. These are termed as the ideal domain labels for the domain invariant feature generator network G, and are represented by the variable $\hat{d}_gen$. 
Hence, the loss for the domain invariant feature generator network $G$, represented by the variable $l_G$, can be calculated by comparing the actual predicted domain labels $\hat{d}$ with the ideal domain labels $\hat{d}_gen$ through an appropriate loss function. As with the domain discriminator network, due to the binary nature of the domain prediction subtask, the Binary Cross Entropy loss function is used to calculate the loss value $l_G$. This can be represented mathematically using the following formula, where the variable $N$ indicates the total number of images in the batch used for training, and the subscript $i$ is used to denote each individual image:

\begin{equation}
\begin{aligned}
l_G&={-{\frac{1}{N}}} \sum_{i=1}^{N} {\; \hat{d}_{gen} log(\hat{d}_i) + \hat{d}_{gen} log(1-\hat{d}_i)} \\
&= {-{\frac{1}{N}}} \sum_{i=1}^{N}{ \; {\frac{1}{2}} log(\hat{d}_i) + {\frac{1}{2}} log(1-\hat{d}_i)} \\
&= {-{\frac{1}{N}}} \sum_{i=1}^{N}{ \; {\frac{1}{2}} \left( log(\hat{d}_i) + log(1-\hat{d}_i) \right)}
\end{aligned}
\end{equation}

%l_G=-1/N ∑_(i=1)^N▒d ̂_gen ×log⁡(d ̂_i )+d ̂_gen×log⁡(1-d ̂_i )
%=-1/N ∑_(i=1)^N▒0.5×log⁡(d ̂_i )+0.5×log⁡(1-d ̂_i )
%=-1/N ∑_(i=1)^N▒0.5×(log⁡(d ̂_i )+log⁡(1-d ̂_i ) )

The generator loss value $l_G$  is used to update the weights of the domain invariant feature generator network $G$, while the domain discriminator loss value $l_D$ is utilized to update the weights of the domain discriminator network $D$. This is done by differentiating the respective loss values with respect to the weights of the corresponding networks, and then multiplying the resulting gradients with an appropriate learning rate before using the resulting values to update the networks themselves. In the following discussions, this particular learning rate is referred to as the domain invariance learning rate, and represented by the variable $\alpha_{DI}$. We implemented DIFL using Tensorflow and the Codes \footnote[1]{\textbf{https://github.com/Sourajit2110/DIFL}} for this project are made publicly available.

\section{Experimental Design}

We now detail the experimental design that was undertaken to develop the final domain invariant feature learning (DIFL) model.

\subsection{Datasets}

\begin{figure*}
  \includegraphics[width=\textwidth]{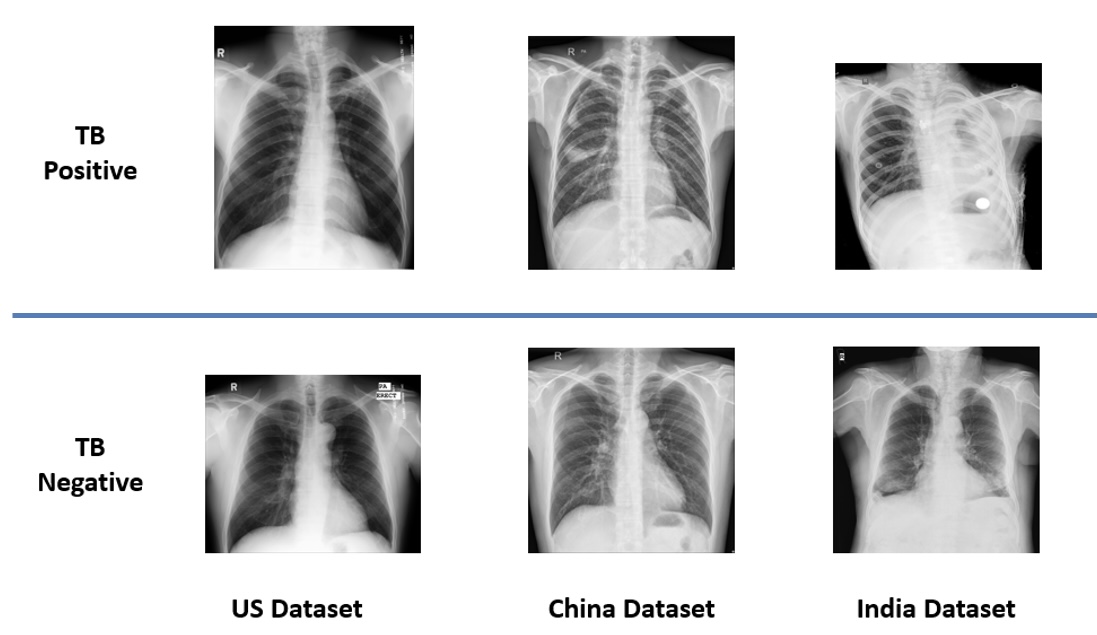}
  \caption{Sample of Chest X-rays from the four datasets.}
  \label{fig3}
\end{figure*}

The datasets used for the purpose of testing and evaluating the models are listed in this section. All the datasets comprise of chest X-ray scans, which were taken for the primary purpose of detecting TB. 

Jaeger et. al \cite{Jaeger_2014} presents two public TB datasets that were used in the purposes of this study. The first dataset consists of chest X-ray scans collected at Shenzhen No.3 People’s Hospital, Guangdong Medical College, Shenzhen, China, and this dataset consists of 326 healthy cases and 336 TB cases, having a total of 662 X-ray images. The second dataset comprises of chest X-ray scans taken through the cooperation of Department of Health and Human Services, Montgomery County, Maryland, USA, and this dataset has a total of 138 X-ray images, of which 80 are healthy cases and the remaining 58 are TB cases. The third dataset was obtained from Chauhan et. al \cite{Chauhan_2014}, which has chest X-ray scans taken from the National Institute of Tuberculosis and Respiratory Diseases, New Delhi, India. This dataset has a total of 176 chest X-ray images, of which 102 are healthy cases and the remaining 74 are TB cases.

Throughout this paper, these respective datasets shall be referred to as the China Dataset, USA Dataset and India Dataset which is based on the country of origin. The datasets that are used for training and testing the models listed in this paper utilize a 80:20 train-test split, both for the source and target domain datasets. Therefore, only 80\% of the dataset is used for training, while the other 20\% of the data is not seen by the model until the final testing stage. Samples images of the 3 datasets, from both the TB positive and TB negative classes, are shown in Figure \ref{fig3}. 

\subsection{Model Development}

In order to have a baseline model for comparison, two non-DIFL models were first developed. Both of these models’ performances were evaluated by using each of the 3 TB datasets as the source domain, and the remaining 2 TB datasets as the target domains. 

The first non-DIFL model was developed using a simple convolutional neural network architecture. Another non-DIFL model was then implemented using the much more complex ResNet50 architecture. The performance metrics of these models on the aforementioned task are detailed in the Results section. Subsequently, the DIFL model was then implemented, as described in the Methodology section. This model was further evaluated in a similar fashion as with the non-DIFL model. Manual hyper-parameter tuning of the DIFL model was performed to determine the classification learning rate and the domain invariance learning rate for the DIFL models. 

\subsection{Evaluation Metrics}

Several measures were used to evaluate the performance of the DIFL model and determine its effectiveness at the task of domain adaptation. These measures are detailed in the following discussion. 

\subsubsection{Accuracy}

Accuracy of label classification is one of the most crucial measures of evaluating the DIFL model, as it can give us a direct idea of how well it can perform in its primary task of TB prediction.

Calculating the accuracy of the DIFL model is straightforward. Input images and their corresponding classification labels $(x,y)$ are sampled from any domain, i.e. either source or target domain. The input images are passed through the domain invariant feature generator network $G$, and to produce the domain invariant features, $G(x)$. These features are then passed through the label classifier network $C$, to produce $C(G(x))$, the predicted classification label, which can also be represented as $\hat{y}$. This predicted label is compared with the true classification label $y$. 

The above process is repeated for all data in the domain, and a confusion matrix is built, consisting of the number of true positives (TP), false positives (FP), true negatives (TN) and false negatives (FN). The accuracy of the DIFL model for this particular domain can then be calculated using to the following formula:

\begin{equation}
    Accuracy = \frac{TP+TN}{TP+TN+FP+FN}
\end{equation}

The resulting accuracy value would be a great indicator of how well the DIFL model is able to correctly classify the input data. However, considering a single performance metric would not be a good enough basis to fully evaluate a model’s performance. Hence, additional metrics are also used to supplement our analysis.

\subsubsection{Sensitivity}

Another metric which can be utilized to analyze the DIFL model is sensitivity. Evaluating the sensitivity can be done in a similar fashion as with accuracy, by first building the confusion matrix. The sensitivity value is then calculated using the following formula:

\begin{equation}
    Sensitivity = \frac{TP}{TP+FN}
\end{equation}

The sensitivity value provides a good measure of how well the model can correctly classify the TB positive instance. Essentially, it represents the proportion of TB positive cases that were correctly predicted as TB positive by the model as well, and hence this value is also known as the true positive rate. 
Apart from sensitivity, the specificity is also a possible metric that can be used to evaluate the DIFL model. Using the confusion matrix, the specificity value is calculated using the following formula:

\begin{equation}
    Specificity = \frac{TN}{TN+FP}
\end{equation}

The specificity value is similar to the sensitivity value, but instead indicates how well the model can correctly classify the TB negative instances, i.e. it represents the proportion of TB negative cases that were correctly predicted as TB negative by the model. Thus, this value is also known as the true negative rate. 

\subsubsection{ROC – AUC Measure}

The Receiver Operating Characteristic (ROC) curve is a graphical plot that visualizes the classification ability of a binary classifier model, by varying its discrimination threshold and evaluating its performance on the data. This is done by plotting a graph of its True Positive Rate (Sensitivity) against its False Positive Rate $(1-Specificity)$ as its discrimination threshold is varied from $0$ to $1$. While the ROC curve enables one to visualize a model’s performance, the graph alone does not however provide a concrete method to objectively compare the performance of different models. 

To supplement the analysis of a model’s ROC curve, an additional metric, known as the Area Under Curve (AUC) value, can be calculated and used to judge the performance of the binary classifier model. The AUC value can be determined trivially by taking the area under the ROC curve (this can be obtained by integrating with respect to the ROC curve), and provides a simple but effective way of directly comparing the performances of different models. 

As the True Positive Rate and False Positive Rate ranges from a minimum value 0 to a maximum value of 1, the maximum AUC value possible would be $1 \times 1=1$, which is only obtainable by a model that is able to perform perfect binary classification. A random binary classifier is expected to achieve an AUC value of 0.5, as it can perform correct predictions about approximately only half of the time. As the AUC value tends closer to $1$, we can infer that the model is able to better classify the instances of the data, and hence is indicative of better model performance. 

\section{Results}

\begin{figure*}
  \includegraphics[width=\textwidth]{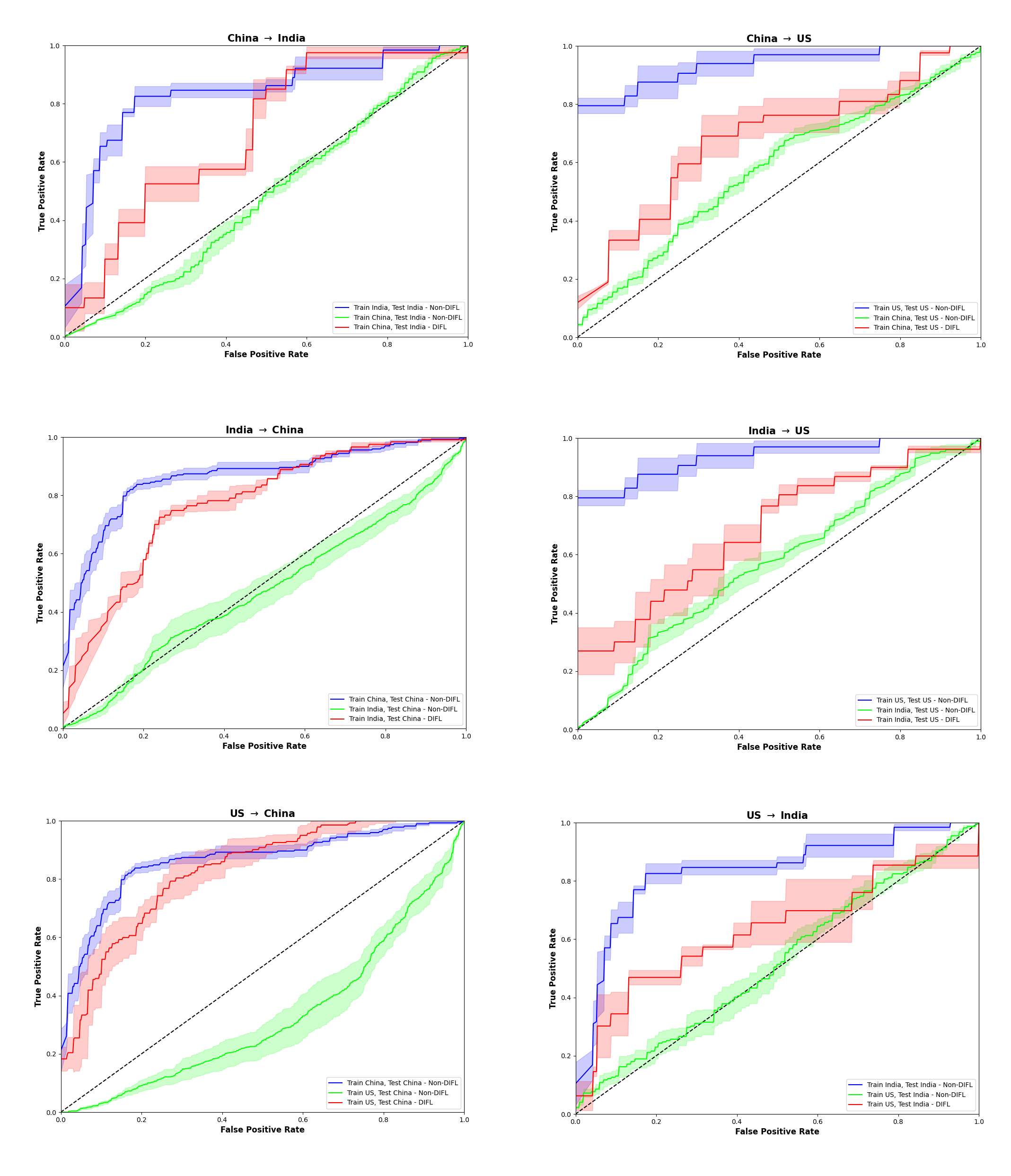}
  \caption{Combined Results: Quantitative Comparison of Performance Exhibited by the DIFL Model across Different Source - Target Domain Combinations.}
  \label{fig4}
\end{figure*}

In this section, the obtained results from the aforementioned testing will be detailed and discussed. 

\subsection{Regional Datasets}

In the first round of experiments, the three regional datasets were utilized, with each of the China, India and US datasets being utilized as the source dataset, while one of the remaining datasets was used as the target dataset. There are six such possible combinations among by using these three regional datasets, and a total of 10 trials were conducted for each of these 6 combinations. The obtained results were then averaged, and are presented below. For the sake of simplifying labels, the $X \rightarrow Y$ nomenclature is used to define the source and target datasets, where X is the source dataset, and Y is the target dataset. 

In each trial per combination, three different types of models are trained and evaluated accordingly: the first model is a non-DIFL model, which is \textbf{trained on the source dataset}, and then \textbf{tested on the target dataset}. This model, as expected, does not achieve good performance measures, due to the presence of domain shift. As such, the performance scores achieved by this model are used as a baseline, and hence this model is termed as the lower baseline model. 

The second model is a non-DIFL model, wherein the model is directly \textbf{trained on the target dataset}, and consequently \textbf{tested on the target dataset}. As it is being directly trained on the data upon which it is also tested, this model is expected to perform well. The performance scores from this model provide an “upper bound” to which the results from the experimental DIFL model can be compared against, and thus this model is termed as the upper baseline model. 

The final model is the DIFL model, which is \textbf{trained on the source dataset}, and \textbf{tested on the target dataset} utilizing the DIFL algorithm. The performance of this model can be evaluated by comparing it against the previously mentioned lower and upper baseline models.

It is also important to mention that the upper baseline non-DIFL, lower baseline non-DIFL and DIFL models contain \textbf{ResNet-50 as the backbone architecture} and using the same architecture among these three models helps us make a justified comparison of their performance. Because, the resultant difference in performance of these three models then is engendered from the training procedure, dataset selection mechanism and incorporation of DIFL method in the pipeline and not the CNN architecture itself.

\subsection{Accuracy Scores}

\begin{table}[htb]
\caption{China $\rightarrow$ India Accuracy Scores}
\setlength{\tabcolsep}{3pt}
\centering
\resizebox{2in}{!}{%
\begin{tabular}{|l|l|}
\hline
\textbf{Type of Model} & \textbf{Accuracy}  \\ \hline
Lower Baseline Model   & $0.47 \pm 0.01$ \\ \hline
DIFL Model             & $0.68 \pm 0.02$ \\ \hline
Upper Baseline Model   & $0.84 \pm 0.01$ \\ \hline
\end{tabular}%
}
\label{tab1}
\end{table}

\begin{table}[htb]
\caption{China $\rightarrow$ USA Accuracy Scores}
\setlength{\tabcolsep}{3pt}
\centering
\resizebox{2in}{!}{%
\begin{tabular}{|l|l|}
\hline
\textbf{Type of Model} & \textbf{Accuracy}  \\ \hline
Lower Baseline Model   & $0.59 \pm 0.01$ \\ \hline
DIFL Model             & $0.71 \pm 0.01$ \\ \hline
Upper Baseline Model   & $0.93 \pm 0.02$ \\ \hline
\end{tabular}%
}
\label{tab2}
\end{table}

\begin{table}[htb]
\caption{India $\rightarrow$ China Accuracy Scores}
\setlength{\tabcolsep}{3pt}
\centering
\resizebox{2in}{!}{%
\begin{tabular}{|l|l|}
\hline
\textbf{Type of Model} & \textbf{Accuracy}  \\ \hline
Lower Baseline Model   & $0.50 \pm 0.02$ \\ \hline
DIFL Model             & $0.73 \pm 0.02$ \\ \hline
Upper Baseline Model   & $0.83 \pm 0.01$ \\ \hline
\end{tabular}%
}
\label{tab3}
\end{table}

\begin{table}[htb]
\caption{India $\rightarrow$ USA Accuracy Scores}
\setlength{\tabcolsep}{3pt}
\centering
\resizebox{2in}{!}{%
\begin{tabular}{|l|l|}
\hline
\textbf{Type of Model} & \textbf{Accuracy}  \\ \hline
Lower Baseline Model   & $0.49 \pm 0.03$ \\ \hline
DIFL Model             & $0.63 \pm 0.04$ \\ \hline
Upper Baseline Model   & $0.93 \pm 0.02$ \\ \hline
\end{tabular}%
}
\label{tab4}
\end{table}

\begin{table}[htb]
\caption{USA $\rightarrow$ China Accuracy Scores}
\setlength{\tabcolsep}{3pt}
\centering
\resizebox{2in}{!}{%
\begin{tabular}{|l|l|}
\hline
\textbf{Type of Model} & \textbf{Accuracy}  \\ \hline
Lower Baseline Model   & $0.45 \pm 0.01$ \\ \hline
DIFL Model             & $0.80 \pm 0.01$ \\ \hline
Upper Baseline Model   & $0.82 \pm 0.01$ \\ \hline
\end{tabular}%
}
\label{tab5}
\end{table}

\begin{table}[htb]
\caption{USA $\rightarrow$ India Accuracy Scores}
\setlength{\tabcolsep}{3pt}
\centering
\resizebox{2in}{!}{%
\begin{tabular}{|l|l|}
\hline
\textbf{Type of Model} & \textbf{Accuracy}  \\ \hline
Lower Baseline Model   & $0.52 \pm 0.02$ \\ \hline
DIFL Model             & $0.63 \pm 0.02$ \\ \hline
Upper Baseline Model   & $0.83 \pm 0.01$ \\ \hline
\end{tabular}%
}
\label{tab6}
\end{table}

The accuracy scores of all three of these models, for each of the six possible combinations amongst the regional datasets, are detailed in Table 1-6.

It is observed that in all the six combinations, the non-DIFL lower baseline model is only able to achieve accuracy scores of around $0.5$, signifying that these models are not performing any better than random guessing. The upper baseline model, as expected, achieves high accuracy scores in the region of $0.8-0.9$.

Looking at the DIFL models, it is observed that they achieve significantly higher accuracy scores than their respective lower baseline models, by approximately $0.2$. While it is not able to achieve accuracy scores as high as the upper baseline models, it does come close in one combination, particularly the US → China case, wherein the DIFL model achieves an accuracy score of $0.80$ (which is a great increase from the lower baseline model’s accuracy of $0.45$), while the upper baseline model achieves an accuracy score of $0.82$. 

Hence, while the DIFL model is not able to outperform the upper baseline model, it can be said that in some cases, it has the capability to achieve a similar performance as that of the upper baseline model. 

\subsection{Presence of Domain Shift}

Upon conducting analysis of the obtained results in the previous sections, the presence of domain shift is confirmed, and its consequence in preventing conventional models to fail when tested on datasets other than the one they have been trained on is evident.  

The lower baseline model, when trained on dataset $X$ and tested on dataset $Y$, was unable to perform as well as the upper baseline model, which was trained on dataset $Y$ and tested on dataset $Y$, even though the classification task is identical, and the image data from datasets $X$ and $Y$ are largely similar to the human eye. This provides concrete evidence that the drop in performance when testing on other datasets does not arise due to potential problems in the other datasets, but rather due to the presence of domain shift, as the non-DIFL model architectures used in both the lower and upper baseline models are identical.

It is also shown that the domain shift present between the source and target datasets can be mitigated by utilizing the proposed DIFL approach. By producing generalized features of the data from both the source and target domain first, before using those features for the classification task at hand, the DIFL algorithm enables the model to perform significantly better when tested on the target dataset than the lower baseline model, which is indicative of the fact that the DIFL model is able to generalize better across these domains.

\subsection{Discussion of Disease Presentation}

One interesting observation from the conducted experiments was the case of using the US dataset as the source dataset, and the Chinese dataset as the target dataset. This particular combination saw the highest improvement in performance from the lower baseline model, and the closest performance in comparison to the upper baseline model. 

A sample of images was qualitatively inspected by a board certified chest radiologist.  Upon further analysis of the chest X-ray imagery from both datasets, this particular observation can be attributed to the differences in disease presentation in the datasets.

As TB is a progressive disease, its effects on the lungs can be classified into two main broad categories; more severe TB (significant damage to the lungs), and less severe TB (minimal damage to the lungs).	As such, images which are classified as TB positive in the datasets could belong to either of these categories. However, it is important to note that the disease presentation of TB in these two categories may be different, and may not contain the same indications on chest X-rays. However, they are not exclusive; particularly, features of less severe TB cases are likely to be present in more severe TB cases, but the converse is not true. 
	
The US dataset contains chest X-ray imagery with less severe manifestations of TB, while the Chinese dataset comprises mostly of chest X-ray imagery with more severe manifestations of TB. As such, when the US dataset is used as the source (generalizing using features from the US dataset), the DIFL model is also able to perform relatively well on the Chinese dataset, as features from less severe manifestations of TB can generalize well to detect more severe manifestations of TB. 

However, the reverse is not true. Features from more severe manifestations of TB do not generalize well to detect the less severe manifestations of TB. As such, this is seen and confirmed with the observations from using the Chinese dataset as the source, and the US dataset as the target; this particular combination does not perform as well as the aforementioned inverse combination.

As such, it is crucial to consider the nature of the datasets, particularly in the context of screening of TB from chest X-ray images, when employing the DIFL approach. Due to the complex nature of TB, factors such as disease presentation may affect the effectiveness of the DIFL approach, and it is important to evaluate these factors before deciding on the source and target datasets. 

\section{Conclusion}
We present an analysis of the cross-domain performance of 
ResNet50 for tuberculsis classification from three geographically distributed data sources, as well as a novel DIFL method for unsupervised domain adaptation in order to mitigate the effects of domain shift on out-of-domain performance.

We demonstrate that the baseline ResNet50 non-DIFL models are unable to generalize to international TB datasets for which they were not trained on, even if the actual classification task (TB screening) is identical, and the x-ray images look very similar to a human Radiologist. This is largely due to the presence of “domain shift”, which causes a non-DIFL model trained on a source dataset to underperform when tested on other target datasets. Utilizing a DIFL approach mitigates this problem by using unlabeled data from the target dataset to produce more generalized features from the source and target dataset, upon which the classification task is then conducted. As such, the DIFL approach enables us to perform classification tasks on the target dataset, even when the data from the target dataset is unlabeled.

Additionally, while it is observed that the DIFL model performs significantly better than non-DIFL models, it does not perform as well as the model which is directly trained on the target dataset in most cases. However, there are circumstances in which the DIFL model performs comparably well to a model which is directly trained on the target dataset.  We believe that this variation in generalizability is largely dependent on the disease presentation of the datasets. Although future work is necessary to confirm this hypothesis. 

In the context of medical imaging and TB, it is observed that the chest X-ray images from the Chinese dataset have disease presentation which contains more severe appearance of TB in the opinion of a board certified radiologist, as compared to chest X-ray images from the US dataset have disease presentation which has a less severe appearance. By using the US dataset as the source dataset, and the Chinese dataset as the target dataset, the DIFL model achieved comparable performance in comparison with the model that was directly trained on the Chinese dataset. 

%Thus, when implementing DIFL algorithms in the context of TB screening with chest X-ray images, choosing the right source dataset is a critical choice. Source datasets with less severe manifestations of TB enables the DIFL model to generalize better, while the inverse (source datasets with more severe manifestations of TB), does not seem to be as effective. However, all DIFL models significantly outperform their non-DIFL counterparts when trained on the source dataset and tested on the target dataset.

\section{Future Work}

The field of DIFL, and domain adaptation in general, has great potential for many purposes, in particular ensuring that algorithms exhibit tailored performance to a given hospital institution.  However, further algorithmic refinement is necessary to demonstrate the effectiveness of this method on a consistent basis, and retraining algorithms to a particular hospital demographic brings up additional regulatory challenges that are continuing to evolve.  This section discusses a few of the possible extensions through which one could continue the work that has been done in this paper. 

The main facet of this experiment which could be improved to achieve better results would be hyperparameter tuning, particularly that of the classification step learning rate and the domain invariance step learning rate. Deciding the appropriate classification step and domain invariance step learning rates was performed through the use of manual tuning, but this is a time intensive process.  We anticipate that it is possible to improve the loss function of the DIFL algorithm such as to maximize directly for the expected target accuracy rather than having two separate objectives to optimize (source accuracy and generalizability).  In this way, it may be possible to eliminate this hyperparameter tuning step. 

%If the classification step learning rate to domain invariance step learning rate is set to be too high, then the model would be able to perform well on the source dataset, but unable to perform equivalently well on the target dataset. Conversely, if the classification step learning rate to domain invariance step learning rate is set to be too low, then the model would be able to generalize well to both the source and target domains, but would not be able to perform classification well on both these domains. As such, striking the right balance between classification step learning rate and domain invariance step learning rate could be a rather time-consuming and challenging process.

%However, instead of using trial and error to determine the best set of values for these hyperparameters, one could take a more systematic approach, by using linear algebra to solve for the ideal values of these hyperparameters, by including them along with the mathematical functions of the neural networks. While this might make the DIFL neural network architecture more complex, it could produce better results as it would ensure that the DIFL model is achieving the right balance of generalization and learning classification features.

Apart from utilizing DIFL algorithms for classification of TB from chest X-rays, one could also attempt to implement similar DIFL algorithms for classification of other diseases from other types of medical imaging, e.g. chest X-rays, mammograms, etc. The effectiveness of DIFL algorithms on other disease presentations have been largely unexplored, and these could be an area of interest for anyone seeking to expand upon the subject of this paper. 

Additionally, another aspect which could be tested would be evaluating the effectiveness of the DIFL algorithm when multiple source datasets or multiple target datasets are involved. While it is shown in this paper that the DIFL algorithm, in the context of TB screening from chest X-rays, can generalize and mitigate the domain shift between a single source dataset and a single target dataset, its effectiveness in dealing with multiple source or target datasets is unknown. The DIFL algorithm could be modified slightly to make this change possible, and experiments could be conducted to assess how using multiple source or target datasets affects the end result.

%\appendices

\section*{Acknowledgment}

This work was supported in part by the NSF Center for Advanced Real-time Analytics Grant 1747724 and in part by Google Foundation.

%\bibliographystyle{unsrt}
%\bibliography{main}

%\nocite{*}

\begin{IEEEbiography}[{\includegraphics[width=1in,height=1in,clip,keepaspectratio]{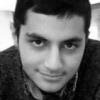}}]{Nishanjan Ravin} is currently working as a Software Development Engineer at Amazon in New York City Metropolitan Area. He received his B.Tech. degree in Mechanical Engineering from National Institute of Technology Tiruchirappalli, India, in 2019. Afterwords, he received his M.S. in Computer Science in 2021 from University of Maryland Baltimore County. His research interests are in Artificial Intelligence, Computer Vision and Machine Learning.  At University of Maryland Baltimore County, he worked on research related to the use of Domain Adaptation in order to address potential biases in deep Tuberculosis screening algorithms due to domain shift.  At Amazon, Nishanjan Ravin currently works in Research and Development in innovative technologies.
\end{IEEEbiography}

\begin{IEEEbiography}[{\includegraphics[width=1in,height=1.30in,clip,keepaspectratio]{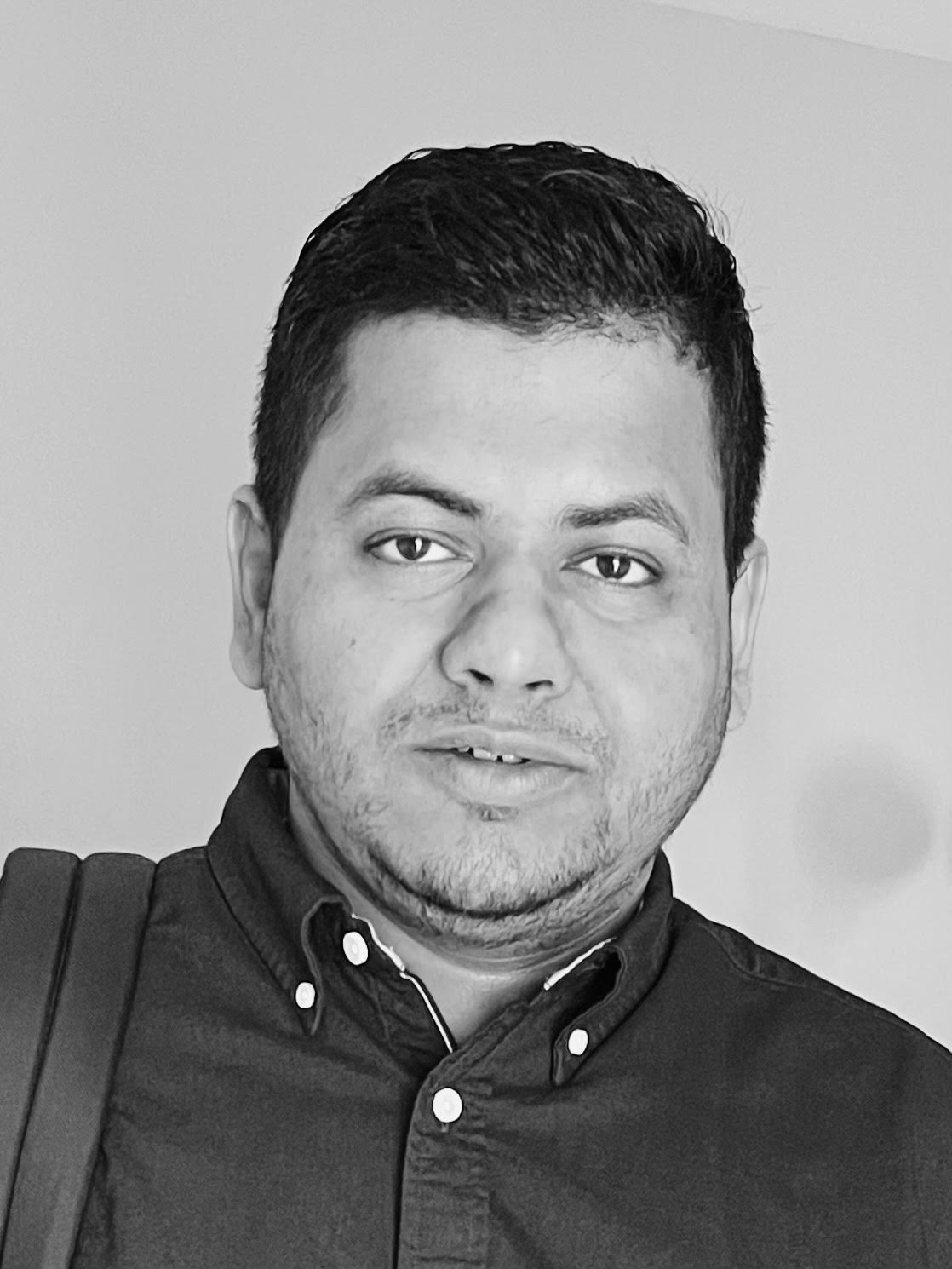}}]{Sourajit Saha} is currently pursuing his PhD degree in Computer Science at University of Maryland Baltimore County, MD, USA. His primary focus of research is Computer Vision and Natural Language Processing with intersection in Machine Learning and Stochastic Optimization. He received his B.S. degree in Computer Science from BRAC University, Dhaka, Bangladesh, in 2017. He worked in Research and Academia in the School of Engineering and Computer Science at Independent University Bangladesh, Dhaka, Bangladesh, from 2018 to 2020. Furthermore, he worked in Computer Vision and Deep Learning as a Research Assistant at Center for Cognitive Skills Enhancement, Dhaka, Bangladesh, from 2018 to 2019. He also served as a Research Collaborator at FAB Lab IUB, Dhaka, Bangladesh, from 2018 to 2020.
\end{IEEEbiography}

\begin{IEEEbiography}[{\includegraphics[width=1in,height=1.30in,clip,keepaspectratio]{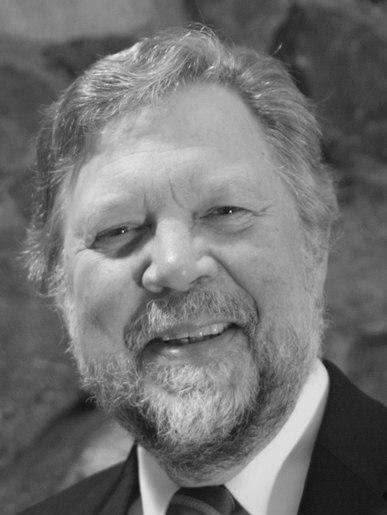}}]{Alan Schweitzer} received his B.S. and M.E.E. degree from Cornell University. He is currently working as the Technical Director of Informatics at RAD-AID International. He provides technical direction for RAD-AID Informatics group, including system architecture, vendor technical liaison, and management of Radiology I.T. initiatives at RAD-AID partner sites. Previously he worked as a CTO at Radiology Consulting Group (RCG), Massachusetts General Hospital. He provided consultation services for all areas of Radiology Information Technology including planning, procurement, vendor selection, and implementation of PACS, Radiology Information Systems, and enterprise medical imaging solutions. Assisted more than 50 clients worldwide with a variety of imaging-related information technology initiatives. Prior to that he was the CTO at Massachusetts General Imaging Business Development Group. He developed and implemented I.T. infrastructure for MGH Imaging’s Teleradiology service. Mr. Schweitzer has presented at the AHRA, SIIMS, and HIMSS conferences, as well as publishing several articles in the areas of speech recognition, storage strategies, PACS vendor evaluations, and imaging in the developing world.
\end{IEEEbiography}

\begin{IEEEbiography}[{\includegraphics[width=1in,height=1.30in,clip,keepaspectratio]{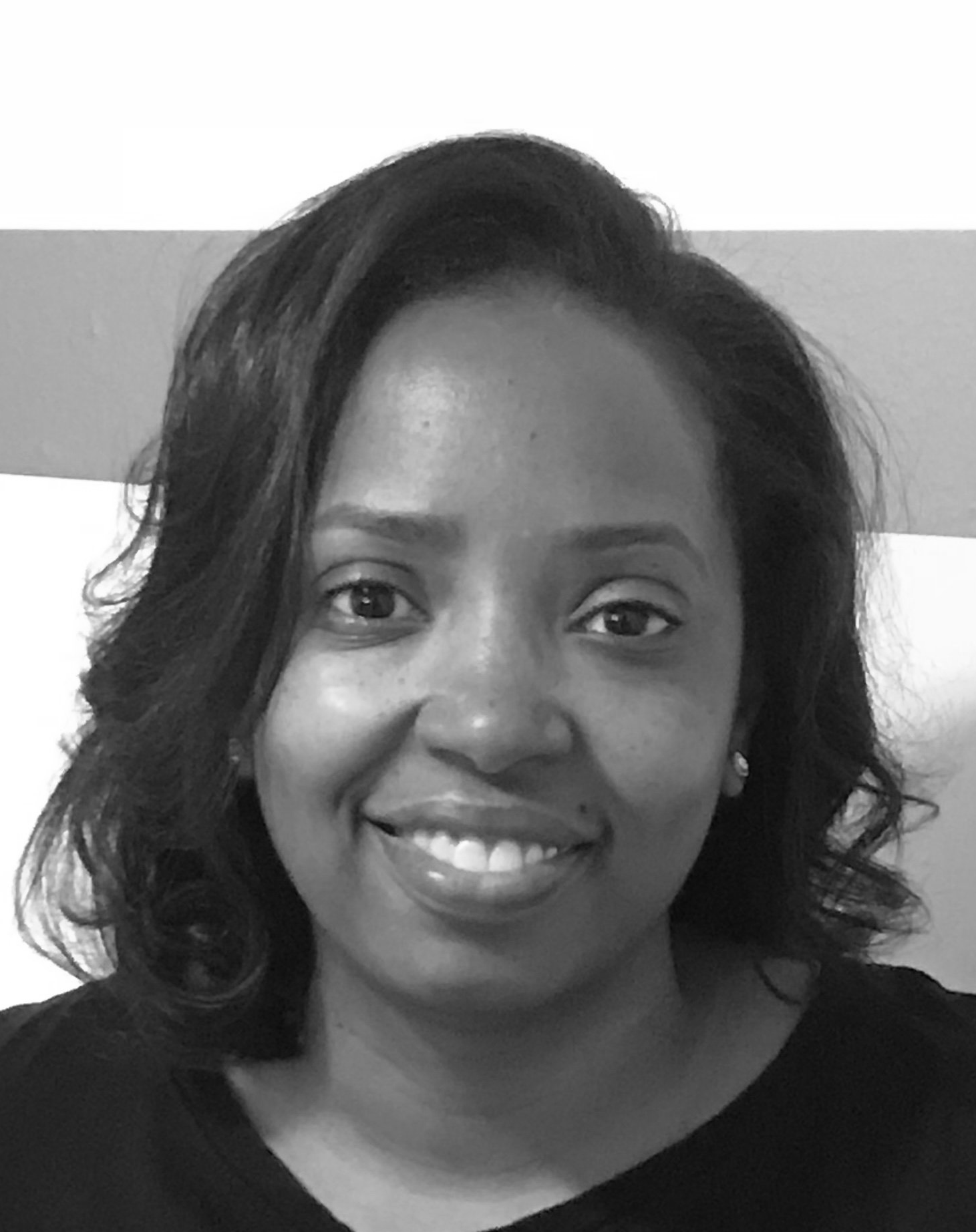}}]{Ameena Elahi} received her B.S, R.T degree in Health Administration and Medical Imaging from Drexel University in 2003 and M.P.A in Public Administration from Keller in 2011. Ameena Elahi is a Senior Technical Analyst on the Medical Imaging Management (MIM) Team at Penn Medicine's Information Services Department. As Co-Chair of the Clinical Imaging Artificial Intelligence (AI) Program, her duties include the product management and implementation of AI technology for research and evaluation.  Ameena worked with RAD-AID International in Nigeria during a Friendship PACS installation in 2019 and has since joined The RAD-AID leadership team as the Informatics Operations Director to pursue her passion for global outreach and improving healthcare for all. Ameena was also elected to the Society of Imaging Informatics in Medicine (SIIM) in 2021 for a three-year term.  Previously, she held roles as a diagnostic and Interventional Radiologic Technologist at The University of Pennsylvania Hospital.
\end{IEEEbiography}

\begin{IEEEbiography}[{\includegraphics[width=1in,height=1.30in,clip,keepaspectratio]{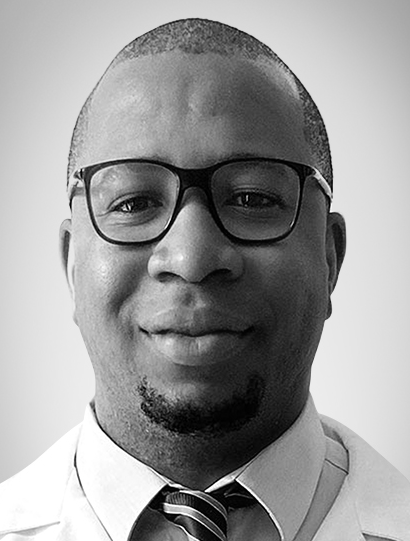}}]{Dr. Farouk Dako} is an assistant professor of radiology in the cardiothoracic imaging division, scholar in the Center for Global Health and senior fellow in the Leonard Davis Institute of Health Economics at the University of Pennsylvania Perelman School of Medicine. He is also the Director for RAD-AID International Nigeria program and has led this program for over 5 years. His research focuses on population and global health, specifically the role radiology can play in advancing health equity. He is interested in the utilization of big data to improve health outcomes of traditionally underserved populations. Dr. Dako attended medical school at St. George’s University. He obtained a Master of Public Health from the Johns Hopkins Bloomberg School of Public Health. He completed a surgical internship at the Mayo Clinic and diagnostic radiology residency at Temple University. He then completed fellowships in cardiothoracic radiology and imaging informatics at the University of Maryland. 
\end{IEEEbiography}

\begin{IEEEbiography}[{\includegraphics[width=1in,height=1.30in,clip,keepaspectratio]{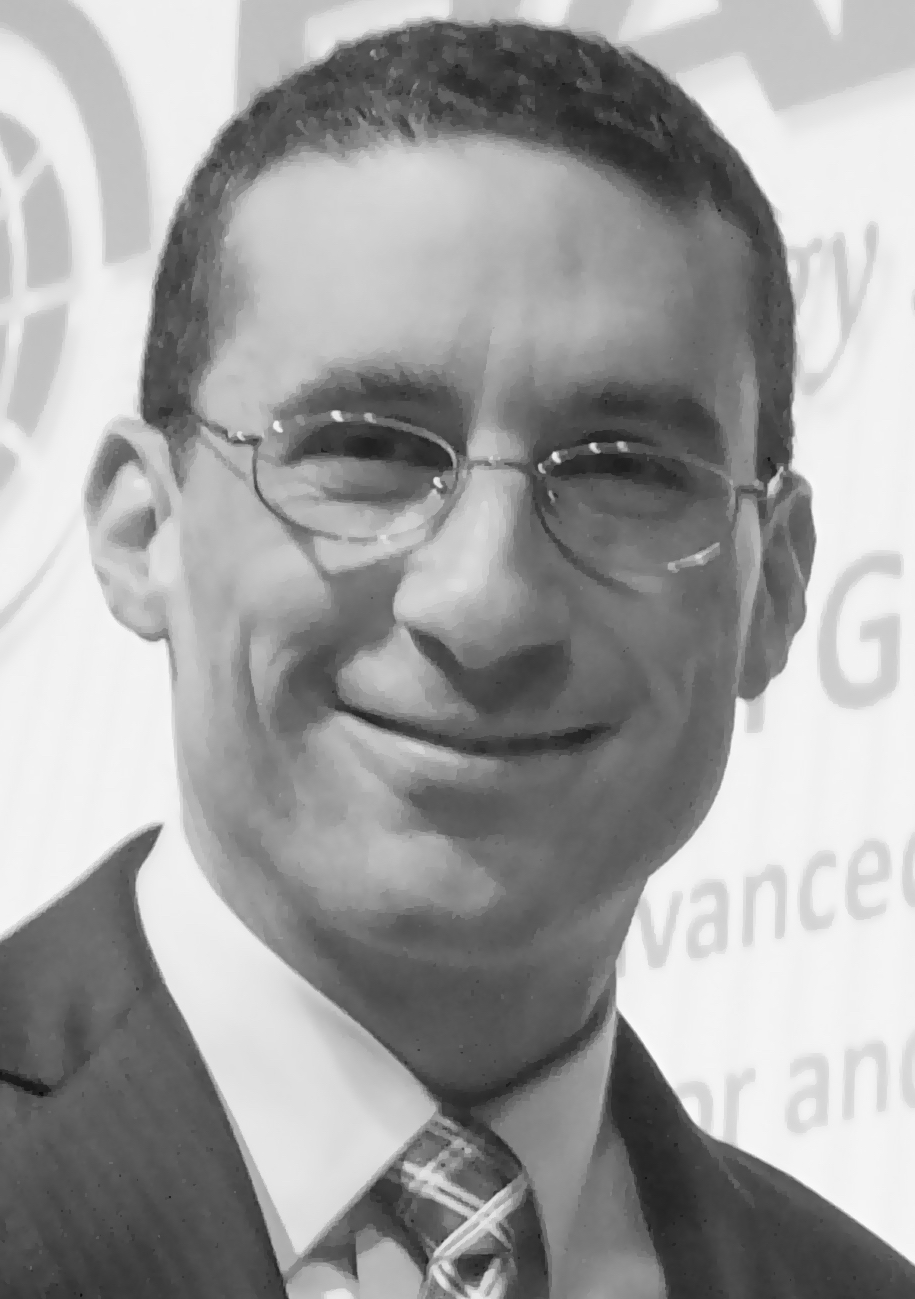}}]{Dr. Dan Mollura} is the Founder, President and CEO of RAD-AID International, a nonprofit 501c3 organization for increasing health and radiology medical care to underserved and resource-poor communities. Dr. Mollura received his undergraduate BA in Government/International Relations from Cornell University (1994), medical degree (MD) in 2003 from the Johns Hopkins University School of Medicine, and completed his internal medicine training (2004), diagnostic radiology residency (2008) and nuclear medicine/molecular imaging fellowship (2009) at the Johns Hopkins Hospital.  Based on his background as a Goldman Sachs Financial Analyst (1994-1996), premedical/scientific research training at Columbia University (1996-1999) and prior founding of three other successful start-ups in the media, technology, and public sectors, Dr. Mollura founded RAD-AID in 2008 to become a global nonprofit organization with nearly 15,000 members serving over 85 hospitals in 38 countries.  Dr. Mollura served for 10 years on the radiology clinical and research faculty of National Institutes of Health (NIH) in Bethesda, Maryland (2009-2019). At NIH, in addition to providing clinical patient-care in radiology and nuclear medicine, Dr. Mollura led an artificial intelligence (AI) laboratory in the NIH Clinical Center to develop and study AI for quantitative molecular imaging applications. In 2020, Dr. Mollura became the full-time President/CEO of RAD-AID and continues to build the charitable global efforts of the organization, including the integral contribution of AI, IT, clinical radiology, education, training, infrastrastructure, and equipment to low-resource health facilities in medically underserved communities and low-middle income countries.
\end{IEEEbiography}

\begin{IEEEbiography}[{\includegraphics[width=1in,height=1.30in,clip,keepaspectratio]{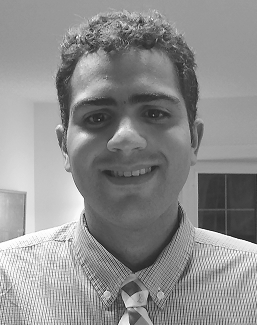}}]{Dr. David Chapman} received his B.S. degree 2006 in computer science from Univ. Maryland Baltimore County (UMBC), his M.S. degree in 2008 from UMBC, and his Ph.D. in 2012 from UMBC. From 2012-2014 he was a postdoctoral fellow at Columbia University, Lamont Doherty Earth Observatory (LDEO). From 2014-2018 he was a design engineer at Oceaneering International Inc. Currently he is an assistant professor with the Dept. of Computer Science at University of Maryland Baltimore County and head of the Vision and Image Processing Algorithms Research Group (VIPAR). VIPAR focuses on computer visions, machine learning, and medical imaging informatics. VIPAR is working in collaboration with University of Maryland School of Medicine, University of California San Francisco, Mercy Medical Center in Baltimore, and RAD-AID informatics on problems related to AI in Radiologicial imaging.  Dr. Chapman's research interests include challenges in the areas of semi-supervised and unsupervised learning as well as improved AI evaluation methodologies toward clinical translation and improved real-world performance.
\end{IEEEbiography}

\EOD

\end{document}